\documentclass[journal=jacsat,manuscript=article]{achemso}

\usepackage[version=3]{mhchem} 
\usepackage{epstopdf}
\usepackage{color}
\usepackage{amsfonts}

\author{Tobias Wassmann}
\author{Ari P. Seitsonen}
\author{A. Marco Saitta}
\author{Michele Lazzeri}
\author{Francesco Mauri}
\email{francesco.mauri@impmc.jussieu.fr}
\affiliation[IMPMC]{IMPMC, Universit\'e Paris 6 et 7, CNRS, IPGP, 140 rue de Lourmel, 75015 Paris, France}

\title[]{Clar's Theory, STM Images, and Geometry of Graphene Nanoribbons}

\makeatletter
\let\section\acs@section
\let\subsection\acs@subsection
\makeatother

\begin{document}
\begin{abstract}
We show that Clar's theory of the aromatic sextet is a simple and powerful tool to predict the stability, the $\pi$-electron distribution, the geometry, the electronic/magnetic structure of graphene nanoribbons with different hydrogen edge terminations. 
We use density functional theory to obtain the equilibrium atomic positions, simulated scanning tunneling microscopy (STM) images, edge energies, band gaps, and edge-induced strains of graphene ribbons that we analyze in terms of Clar formulas.
Based on their Clar representation, we propose a classification scheme for graphene ribbons that groups configurations with similar bond length alternations, STM patterns, and Raman spectra.
Our simulations show how STM images and Raman spectra can be used to identify the type of edge termination.
\end{abstract}

\pagebreak
\section{Introduction}
Since its first characterization in 2004,\cite{Novoselov2004} graphene has developed into a major research topic of its own and 
from the beginning, its unique electron transport properties\cite{Bolotin2008, Morozov2008, Berger2006, Miao2007, Murali2009,Barreiro2009} (For a review, see Ref.~\cite{Neto2009}) have propelled the hope for application in a post-silicon generation of electronic devices.\cite{Novoselov2004,Berger2006}
One of the big advantages graphene has over other potential materials such as carbon nanotubes (CNTs) is that it
can be patterned with lithography methods.\cite{Berger2006}
In this context nanometer sized graphene ribbons come into focus. On one hand, they could serve as conductive interconnects in integrated circuits\cite{Xu2007} and on the other hand as channel material in field effect transistors.\cite{Chen2007, Li2008, Wang2008PRL} 
To achieve sufficient on-off ratios in semiconducting devices, however, the electronic band gap has to be large enough, and, as a consequence, lateral dimensions below 10 nm are required since the band gap is inversely proportional to the width.\cite{Ezawa2006, Barone2006, Son2006b, Yang2007, Han2007, Li2008}.
At this scale, the edge chemistry and geometry determine the electronic properties.\cite{Nakada1996, Gunlycke2007,Gunlycke2007b} 
For instance, single-hydrogen-terminated zigzag ribbons are predicted to feature a spin-polarized edge state whose order is ferromagnetic along the ribbon and antiferromagnetic across the ribbon.\cite{Yamashiro2003, Pisani2007} 
Under high external fields, this state could turn the zigzag ribbons into half-metals, opening interesting perspectives for application in spintronics.\cite{Son2006, Yazyev2008} 
Magnetic ground states were also found for other graphene nanostructures.\cite{Wang2007,Yazyev2008NL,Fernandez-Rossier2007,Hod2008,Wang2009}
In single-hydrogen-terminated armchair ribbons, however, this phenomena is absent.
Despite the broad range of production techniques for graphene ribbons\cite{Han2007,Chen2007,Tapaszto2008,Bai2009,Yang2008,Delgado2008,Li2008,Wang2008,Kosynkin2009,Jiao2009,Datta2008,Ci2008,Campos2009,Wei2009,Gao2009} real control over the edge geometry and termination in them has not been achieved yet, and no atomic characterization either.
For other sp$^2$ bonded materials such as polycyclic aromatic hydrocarbons (PAHs), Clar's theory of the aromatic sextet\cite{Clar1972} has proven to be an intuitive, yet adequate model for the edge-induced $\pi$-electron distribution\cite{Gutman2004} and can account for many properties of PAHs.\cite{Coulson1952,Krygowski2001,Watson2001}
It has also been applied to graphene-related systems,\cite{Watson2001,Balaban2009} carbon nanotubes,\cite{Ormsby2004,Baldoni2007,Baldoni2009} and to ribbons\cite{Mauri2008,Baldoni2008} to some extent.

In this article, we analyze a set of selected, hydrogen-terminated graphene ribbons. 
We suggest a classification scheme that groups ribbons which have similar representations according to Clar's theory.
In an analysis of simulated scanning tunneling microscopy (STM) images and relaxed density functional theory (DFT) coordinates
we show that the Clar formulas correctly account
for the $\pi$-electron distribution, bond lengths, and hexagon areas.
Furthermore, we demonstrate how magnetic edge states can be explained---even quantitatively---with Clar's theory. 
Finally, we present calculations on the width dependence of the electronic band gap, the edge energy, and the edge-induced strain.

\section{Considered systems}\label{sec:notation}
\begin{figure}
	\begin{flushleft}
	\includegraphics[scale=1]{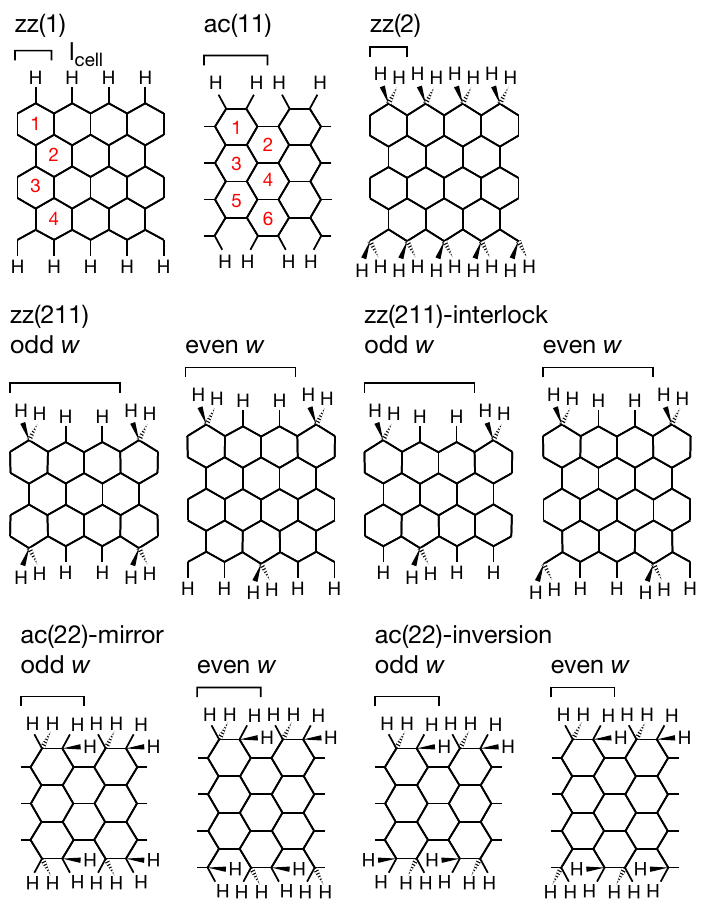}
  	\caption{The ribbon configurations considered in this study with their corresponding names. 
	Periodic boundary conditions were applied to a supercell setup with length $l_{\textrm{cell}}$,  indicated by the bar above each structure. 
	Note that only the $\sigma$-bond network between the carbon atoms is shown. Solid wedges represent bonds pointing out of the plane, towards the reader, dashed ones represent bonds pointing into the plane, away from the reader. The red numbers inscribed in the first two diagrams indicate how we defined the width index $w$ of the ribbons.}
	\label{fig:notation}
	\end{flushleft}
\end{figure}

The graphene ribbons analyzed in this study were the four hydrogen-terminated edge configurations identified in Ref.~\cite{Mauri2008} to be the thermodynamically most stable ones, zz(1), ac(11), zz(211), and ac(22) (see \ref{fig:notation}). 
In addition to that, we included also ribbons with double-hydrogenated zigzag edges, zz(2). 
Note that for the zz(211) edge termination there are two different ribbon configurations, denoted as zz(211) and zz(211)-interlock, depending on the relative position of the double-hydrogenated sites on each edge (second row in \ref{fig:notation}).
In fact, there are also two different geometries for the ac(22) ribbons. 
These were called ac(22)-mirror and ac(22)-inversion depending on whether the spatial arrangement of the hydrogen atoms on the opposite edges is mirror or inversion symmetric with respect to the centre of the ribbon (third row in \ref{fig:notation}). 

Besides the notation, \ref{fig:notation} shows the lattice constant of the supercells, $l_{\textrm{cell}}$, and the definition of the width index $w$, i.e.\ the number of hexagons in the lateral direction, that we used in this study to characterize the width of the ribbons. Calculations were performed for width indices in the range of $w=1,\ldots ,13\, (22)$ for zigzag (armchair) ribbons, translating into widths from $4.3$ to $30.7\,\textrm{\AA}$ between the outermost atoms. 

\section{Computational methods}
We performed first principles calculations with the {\sc PWSCF} code of the {Quantum-\sc ESPRESSO} package\cite{Giannozzi2009} to investigate the atomic coordinates, edge energies, band gaps, and edge-induced strains of the graphene nanoribbons.
Plane-wave basis functions, Vanderbilt ultrasoft pseudopotentials,\cite{Vanderbilt1990} and the PBE generalized-gradient approximation (GGA)\cite{Perdew1996} as the exchange-correlation functional were applied.
The Brillouin zone integration was done with a uniform grid of 12 \mbox{k-points} along the periodic direction for the armchair ribbons and of 24 [8] k-points for the zigzag ribbons zz(1) and zz(2) [zz(211) and zz(211)-interlock] respectively. 
The cutoff energy for the wave functions was set to 30 Ry and the one for the charge to 300 Ry. 
In our supercell setup, vacuum distances of $9.5$ and $8.5\, \textrm{\AA}$ separated the ribbons in plane and between planes. 
The lattice constant, $l_{\textrm{cell}}$, along the ribbon was optimized in a separate loop. 
The atomic positions in the cell were allowed to relax until all the forces on the nuclear coordinates were below a threshold of 0.1 mRy/Bohr. 
With the optimized coordinates, additional calculations were performed for the simulation of STM images using norm-conserving pseudopotentials to avoid the wiggles in the tails of the augmentation densities which are inherent in ultrasoft pseudopotentials. 
For these calculations, the wave function cutoff energy was set to 60 Ry, the cutoff for the charge to 240 Ry, the vacuum between the ribbons to $12.7\, \textrm{\AA}$, and the one between planes to $10.6\, \textrm{\AA}$. 
The Brillouin zone sampling consisted of 60 k-points for the armchair ribbons and of 102 [34] k-points for the zigzag ribbons. 

Simulated STM images were obtained using the Tersoff-Hamann approximation.\cite{Tersoff1983, Tersoff1985} 
In this approach, the local tunneling current $I$ between sample and tip is proportional to the sum of the electron density of orbitals in the interval $[\epsilon_{\textrm{F}}+eU, \epsilon_{\textrm{F}}]$ for a negative and $[\epsilon_{\textrm{F}}, \epsilon_{\textrm{F}}+eU]$ for a positive bias voltage $U$, respectively,
\begin{equation}\label{eq:TH}
\textrm{I}(x,y,z=d, U) \propto \sum_{\mu} \vert \Psi_{\mu}(x,y,z=d) \vert^2\, \Bigl[ f(\epsilon_{\textrm{F}}-\epsilon_{\mu})-f(\epsilon_{\textrm{F}}+eU-\epsilon_{\mu})\Bigr] \, .
\end{equation}
Here, $\epsilon_{\textrm{F}}$ denotes the Fermi energy, defined as lying in the middle between valence and conduction band if a gap is present, $e$ stands for the elementary charge ($e>0$), $f$ is the electron occupation function, $f(\epsilon)=1$ for $\epsilon \le 0$ and $f(\epsilon)=0$ for $\epsilon > 0$, $\epsilon_{\mu}$ is the energy of the state $\mu$, and $d$ the sample-tip distance. We used a fixed sample-tip distance of $d=3\, \textrm{\AA}$. 

We recall that in the case of a negative bias voltage $U$ the electrons tunnel from occupied states of the sample to empty states of the tip.
Consequently, in STM measurements at negative bias voltage the occupied states of the sample are probed whereas the empty states are probed at positive bias. 
In graphene ribbons, the states near the Fermi energy belong to the $\pi$-electrons.\cite{Gutman2004}
So, at small bias, the STM signal is dominated by these states and it essentially reflects the distribution of the $\pi$-electrons over the lattice.

To demonstrate the robustness of the DFT-PBE exchange-correlation functional used in this study,
we compared the PBE energy differences among a set of small PAH molecules with published values obtained with the hybrid B3LYP functional\cite{Becke1993,Stephens1994} and with experimental differences in the enthalpy of formation. 
The results are shown in \ref{tab:PBE_vs_B3LYP}.
Notice that the energies of these PAH molecules, which are isomers of the molecular formula C$_{18}$H$_{12}$, vary significantly. 
The reasons for this 
lie in the steric repulsion between the hydrogen atoms and in different numbers of $\pi$-resonances, i.e.\ ways to arrange the $\pi$-electrons among the bonds of the molecules.\cite{Watson2001} 
The results obtained with the PBE functional are very close to the values of the B3LYP calculations. 
Both DFT results agree well with the experimental values in all cases except for benz[a]anthracene $\mathbf{2}$. 
This could hint to a possible problem in the experimental determination of the enthalpy of formation in this case. 
The good agreement between PBE and B3LYP results attests the employed functional a good reliability for calculations on PAH systems such as graphene nanoribbons.
\begin{table}
\begin{tabular}{rccc}\hline\hline
Molecule & PBE & B3LYP/6-31G*~\cite{Herndon1998} & Exp.~\cite{Slayden2001} \\
\hline
Naphthacene $\mathbf{1}$  & 0.40 & 0.44& 0.40 \\
Benz[a]anthracene $\mathbf{2}$  &  0.08 & 0.08& 0.26 \\
Chrysene $\mathbf{3}$  & 0.00 & 0.00 & 0.00 \\
Benzo[c]phenanthrene $\mathbf{4}$ & 0.24 & 0.26& 0.29\\
Triphenylene $\mathbf{5}$  & 0.03 & 0.03 & 0.02 \\
\hline
\end{tabular}
\caption{Energy differences (in eV) between the test molecules depicted in \ref{fig:clar_isomers} as obtained with PBE and B3LYP functionals as well as experimental differences in the enthalpy of formation. Chrysene was chosen as reference.}\label{tab:PBE_vs_B3LYP}
\label{tab:test_cases}
\end{table}
\begin{figure}
	\begin{flushleft}
	\includegraphics[scale=1]{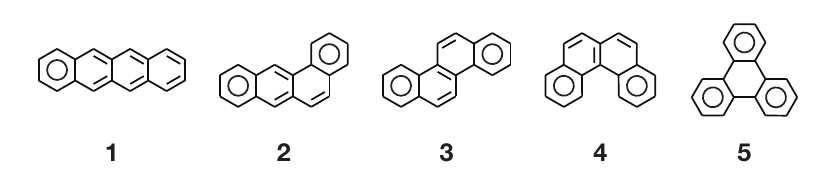}
  	\caption{Clar representations, with implicit hydrogen atoms, of the C$_{18}$H$_{12}$ isomers used as test cases for the evaluation of the PBE functional. $\mathbf{1}$ naphathacene, $\mathbf{2}$ benz[a]anthracene, $\mathbf{3}$ chrysene, $\mathbf{4}$ benzo[c]phenanthrene, and $\mathbf{5}$ triphenylene. 
	}
	\label{fig:clar_isomers}
	\end{flushleft}
\end{figure}

\section{Clar's theory of the aromatic sextet}\label{sec:clar_theory}
\begin{figure}
	\begin{flushleft}
	\includegraphics[scale=1]{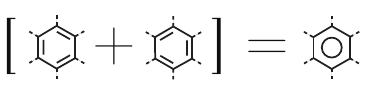}
  	\caption{The Clar sextet represents the delocalization of six $\pi$-electrons resulting form the resonance of two Kekul\'e bond configurations with alternating single and double bonds.}
	\label{fig:clarsextet}
	\end{flushleft}
\end{figure}

In the structure model of PAHs according to Kekul\'e, the four valence electrons of each carbon atom are arranged with electrons of neighboring atoms, forming either single, double, or triple bonds. 
The electronic structure of a system is then the result of a superposition of all possible Kekul\'e bond formulas.
In extension to the Kekul\'e structure model, Clar's theory of the aromatic sextet introduces the \emph{Clar sextet}, a representation for the delocalization of six $\pi$-electrons due to the resonance of two complementary, hexagonal Kekul\'e configurations with alternating single and double bonds (\ref{fig:clarsextet}).
According to Clar's rule,\cite{Clar1972} for a given molecule, the representation with maximal number of Clar sextets, called \emph{Clar formula}, is the most representative one.  
The Clar formula characterizes best the properties of PAHs such as the bond length alternations\cite{Krygowski2001} or the local density of $\pi$-states\cite{Gutman2004} and it is more stable than alternative bond configurations.\cite{Maksic2006} 
This is because a Clar bond configuration with $n$ Clar sextets represents the resonance of $2^n$ Kekul\'e type bond formulas.\cite{Randic2003} So, maximizing the number of Clar sextets unifies a maximal number of resonant Kekul\'e formulas in one single Clar representation. 
Note that, as the bonds sticking out of a Clar sextet are formally single bonds, two adjacent hexagons can never be Clar sextets at the same time. 

\begin{figure}
	\begin{flushleft}
	\includegraphics[scale=1]{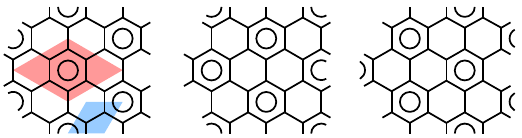}
  	\caption{The three equivalent Clar formulas of graphene. To the left in blue, the primitive unit cell of graphene and in red the unit cell of the $(\sqrt{3}\times\sqrt{3})\textrm{R}30^{\circ}$ superstructure. Notice that only a subsection of the infinite graphene lattice is shown. }
	\label{fig:clar_graphene}
	\end{flushleft}
\end{figure}

Graphene has three equivalent Clar formulas in each of which every third carbon hexagon is a Clar sextet.\cite{Mauri2008,Balaban2009} 
In these Clar formulas, all $\pi$-electrons belong to a Clar sextet and no localized double bonds are present (see \ref{fig:clar_graphene}). 
Such systems that can be represented without localized double bonds are called \emph{all-benzenoid polycyclic aromatic hydrocarbons}. 
They show a particularly high stability, high melting point, and low chemical reactivity.\cite{Mullen2007}
In each of the three Clar formulas of graphene, the pattern of Clar sextets induces a \mbox{$(\sqrt{3}\times\sqrt{3})\textrm{R}30^{\circ}$} superstructure, i.e.\ a reconstruction with a unit cell that is rotated by $30^{\circ}$ and whose sides are scaled by a factor of $\sqrt{3}$ compared to the primitive unit cell (\ref{fig:clar_graphene}).

In PAHs and graphene nanoribbons, however, the presence of edges can lead to a break-down of the prototypical all-benzenoid $\pi$-electron distribution of graphene.\footnote{The circumferential boundary conditions in the case of CNT can have the same effect.} 
Under the premise that all four valence electrons of the carbon atoms are engaged in bonds with neighboring atoms and no dangling bonds or free radicals are present, some edge configurations do not allow the distribution of all $\pi$-electrons into Clar sextets.
In general, the edges of graphene ribbons and PAHs impose a Clar formula consisting of a mix of localized double bonds and Clar sextets. 

In this study, we define graphene ribbons as \emph{(pseudo)-all-benzenoid} if in the Clar formula 
of their unit cell the number of isolated double bonds as a function of the width $w$ is limited by a constant. 
For example, the Clar formulas of the unit cells of ac(11), zz(211), zz(211)-interlock, and ac(22) ribbons 
never feature more than two double bonds, independent of the width of the ribbons (see \ref{fig:1CF}, \ref{fig:2CF}, and \ref{fig:nCF}). 
Notice that in the limit of large ribbons, the density of Clar sextets in these structures approaches that of two-dimensional graphene.
Ribbons in which the number of double bonds in the Clar formula of their unit cell increases with the width, we called \emph{non-benzenoid}. 
Examples of non-benzenoid ribbons are zz(1) and zz(2), as will become apparent in \ref{fig:NB}.

The Clar formula for a given molecule or a periodic cell of a crystal is not necessarily unique. 
We thus further distinguish the (pseudo)-all-benzenoid ribbons into the following three subclasses according to the number of different equivalent Clar formulas they possess:
\begin{enumerate}
\item Ribbons having one, unique Clar formula (called \emph{class 1CF})
\item Ribbons having exactly two Clar formulas (\emph{class 2CF})
\item Ribbons having more than two Clar formulas (\emph{class nCF})
\end{enumerate}
Notice that the definitions \emph{(pseudo)-all-benzenoid} and \emph{non-benzenoid} apply to ribbons independent of their width, meaning all ribbons with a given edge configuration are either (pseudo)-all-benzenoid or non-benzenoid. In contrast to that, the definition of the above subclasses of (pseudo)-all-benzenoid ribbons is defined only for a specific width and may change for different widths.

We want to emphasize that the affiliation of a ribbon configuration to either the (pseudo)-all-benzenoid or the non-benzenoid category is not determined by the edge geometry---zigzag or armchair---alone.
Even though in the present study the class of non-benzenoid ribbons was represented by zigzag structures only, one can easily imagine armchair ribbons which belong to this group. For example the armchair configuration with one of the edge sites being double-hydrogenated while the other is single-hydrogenated, in our notation called ac(21), is also non-benzenoid. 

\section{Results}
\subsection{Geometry and simulated STM images}
\subsubsection{Graphene}
\begin{figure}
	\begin{flushleft}
	\includegraphics[scale=1]{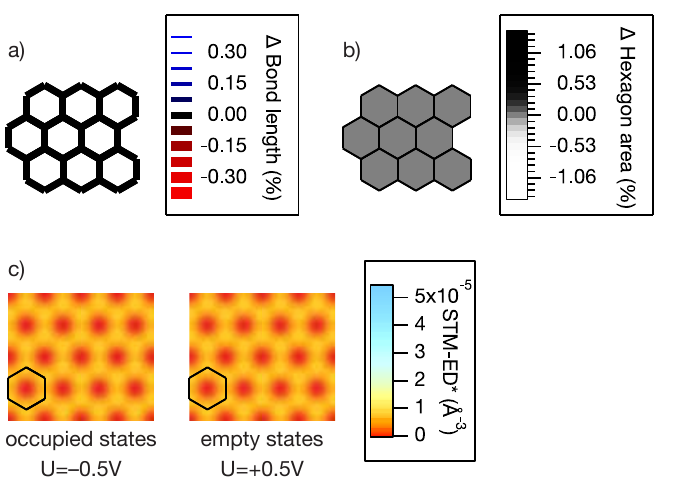}
  	\caption{Bond lengths (a), hexagon areas (b), and simulated STM images (c) of two-dimensional graphene. The scales in (a) and (b) show the deviation form the values $1.427\,\textrm{\AA}$ and $5.293\,\textrm{\AA}^2$, respectively, in percent. $U$ denotes the bias voltage. 
	*~STM relevant electron density.}
	\label{fig:bulk}
	\end{flushleft}
\end{figure}

Before we present the results for the ribbons, it is illustrative to start with infinite two-dimensional graphene.
The superposition of the three equivalent Clar formulas of graphene, depicted in \ref{fig:clar_graphene}, leads to a uniform $\pi$-electron distribution over the crystal with 2/3 $\pi$-electrons per carbon-carbon bond and a uniform bond order of 4/3. As a consequence, all bond lengths and hexagon areas are expected to be equal. 
In \ref{fig:bulk}a and \ref{fig:bulk}b, based on the relaxed DFT coordinates, the carbon-carbon bonds and the hexagon areas 
are color-coded according to their size.
Obviously, all bond lengths and hexagon areas assume the same values, $1.427\,\textrm{\AA}$ and $5.293\,\textrm{\AA}^2$ respectively, and no deviations are present. 
\ref{fig:bulk}c shows simulated STM images of graphene, obtained as absolute values of Eq.\ (\ref{eq:TH}), confirming the uniform distribution of the $\pi$-electrons over the hexagonal carbon lattice. 
The negative and positive bias STM images look very much alike because of the symmetric band structure of graphene around the Fermi energy.

\subsubsection{(Pseudo)-all benzenoid ribbons}
Among the configurations considered here, the ac(11), ac(22), zz(211), and zz(211)-interlock ribbons constitute the (pseudo)-all-benzenoid class. 
While the zz(211)-interlock and zz(211) ribbons have exactly one, respectively two Clar formulas for their unit cell, independent of the width, 
the ac(11) and ac(22) ribbons change 
in a cyclic way between the subclasses 1CF, 2CF, and nCF with increasing width. In these cases, a discussion involving the width index $w$ introduced in \ref{fig:notation} is inevitable. 
Note that the two smallest possible ac(22) ribbons with width indices $w=1$ and $w=2$ are excluded from the discussion. They are too narrow for even one Clar sextet to form, or in other words, they consist of edges only and have no interior region at all.

\begin{figure}
	\includegraphics[scale=1]{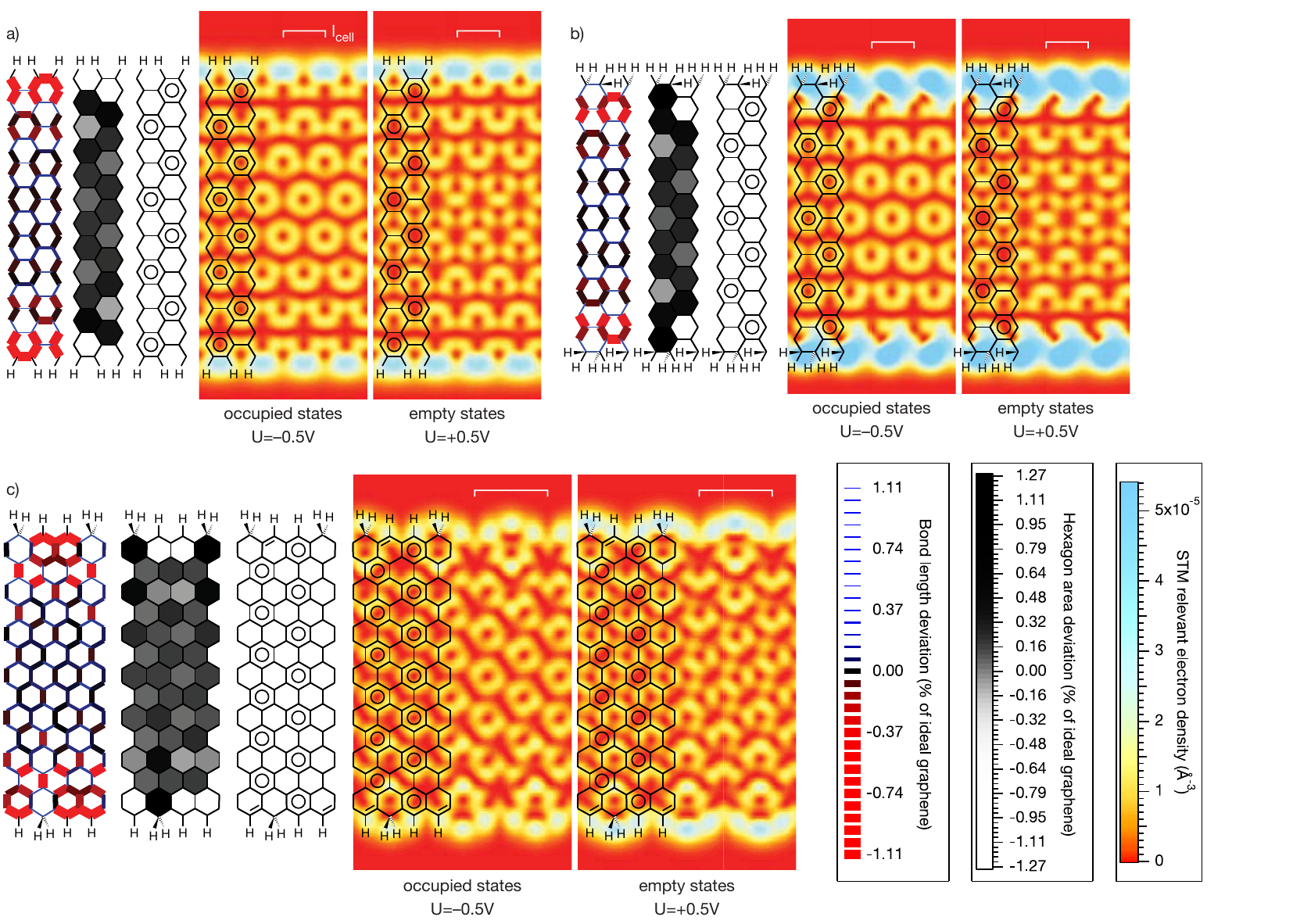}
  	\caption{Graphene ribbons with one unique Clar formula, class 1CF: ac(11) ribbon with width index $w=22$ (a), ac(22) ribbon with width index $w=21$ (b), and zz(211)-interlock ribbon with $w=13$ (c). Subfigures show from left to right: Bond lengths, hexagon areas, Clar formula, negative bias STM image, and positive bias STM image. The scales of the bond lengths and hexagon areas show the deviation form the values of ideal graphene, $1.427\,\textrm{\AA}$ and $5.293\,\textrm{\AA}^2$, respectively, in percent. All scales are valid throughout this section.}
	\label{fig:1CF}
\end{figure}

\paragraph{Subclass 1CF: One unique Clar formula}
The subclass 1CF is made up of ac(11) ribbons with a width index $w=3n+1,\, n$ being integer, ac(22) ribbons with widths of $w=3n$, and zz(211)-interlock ribbons of arbitrary width. They all have only one unique Clar representation with maximal number of Clar sextets for their unit cell as shown in \ref{fig:1CF}.

\subparagraph{Bond lengths.}
The Clar formulas of subclass 1CF configurations show a $(\sqrt{3}\times\sqrt{3})\textrm{R}30^{\circ}$ superstructure of Clar sextets 
like in two-dimensional graphene.
But in contrast to the latter case, here the boundary conditions fix the positions of the Clar sextets.
As the sides of a Clar sextet have a bond order of 3/2,
one would expect the lattice geometries of the subclass 1CF configurations 
to feature a pattern of 
hexagons with smaller sides that are connected by longer bonds. 
The DFT coordinates corroborate this hypothesis (first subfigures in \ref{fig:1CF}a-c). 
Note that the scales for the bond lengths and the hexagon areas in \ref{fig:1CF} use a non-linear partitioning. 
Their range is limited to bit more than $\pm1\,\%$ deviation from the graphene values.
The interesting part of the bond length and hexagon area alternations lie within these boundaries. 
However, at the edges, bond length derivations of up to $6\,\%$ can occur.
In the two armchair configurations ac(11) and ac(22), the bond length alternations are clearly visible and in accordance with the Clar formulas.
Both structures show a pattern of hexagons mainly consisting of dark red or black lines, corresponding to the Clar sextets, and fine blue lines sticking out of them, representing longer single bonds. 
For the zz(211)-interlock configuration, however, the effect is rather weak but still visible in the interior of the ribbon (\ref{fig:1CF}c). 

\subparagraph{Hexagon areas.}
The Clar formulas also imply that the hexagons corresponding to Clar sextets should have a slightly smaller area than the surrounding ones as all the sides of the Clar sextets have bond order 3/2 whereas the \emph{non-Clar} hexagons are made up of three sides with bond order 3/2 and three sides with bond order 1.
The second subfigures in \ref{fig:1CF}a-c show that the DFT coordinates support this. 
The hexagons corresponding to Clar sextets in the Clar formula appear in brighter grey, meaning smaller area than in the surrounding ones.
Again, the pattern of bright hexagons is clearly visible in the two armchair configurations, whereas it is less accentuated in the zz(211)-interlock case but still appears towards the middle of the ribbon.

\subparagraph{Simulated STM images.}
The Clar formulas of subclass 1CF structures indicate that the local density of occupied $\pi$-states should be higher in the hexagons corresponding to Clar sextets than in the neighboring ones. 
The simulated STM images in \ref{fig:1CF} are in perfect agreement with this model. 
Hexagons corresponding to Clar sextets appear as bright rings in simulated STM images with negative bias, reflecting the high local density of occupied $\pi$-states in that area. 
The effect is clear enough in all three configurations and the ring pattern runs through the whole width of the ribbons. 
In positive bias STM images, the Clar sextets correspond to dark regions, reflecting the fact that a large part of the $\pi$-states in this area are occupied and therefore the density of unoccupied states is low. In the bonds sticking out of a Clar sextet, on the other hand, the density of unoccupied states is high as the corresponding electrons are engaged in the formation of a Clar sextet. Hence, these bonds appear bright in positive bias STM images.

\begin{figure}
	\begin{flushleft}
	\includegraphics[scale=1]{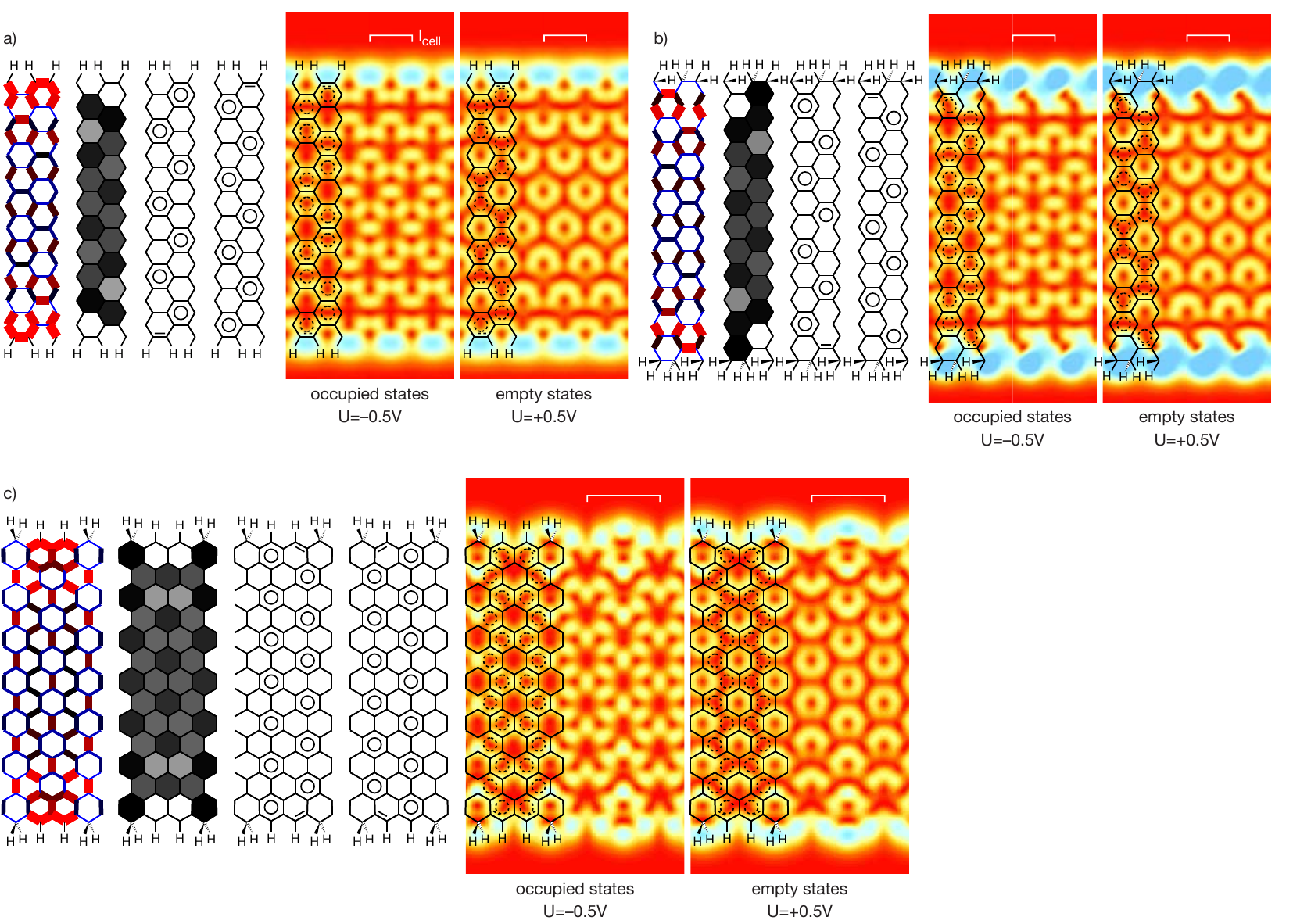}
  	\caption{Graphene ribbons with two equivalent Clar formulas, class 2CF: ac(11) ribbon with width index $w=20$ (a), ac(22) ribbon with width index $w=22$ (b), and zz(211) ribbon with $w=13$ (c). 
	The dashed circles in the lattice on the STM images illustrate the
	superposition of the two Clar formulas.}
	\label{fig:2CF}
	\end{flushleft}
\end{figure}

\paragraph{Subclass 2CF: Two equivalent Clar formulas}
The ac(11) ribbons with a width index of $w=(3n+2),\; n$ being integer, the ac(22) ribbons with widths of $w=(3n+1)$, and all zz(211) ribbons have two equivalent Clar formulas and thus constitute the subclass 2CF. The two Clar formulas of these structures are always connected via symmetry operations (see \ref{fig:2CF}a-c). 

\subparagraph{Bond lengths.}
The subclass 2CF structures are complementary to the subclass 1CF structures.
The superposition of the two equivalent Clar formulas results in a $(\sqrt{3}\times\sqrt{3})\textrm{R}30^{\circ}$ pattern of hexagons which in neither of the two possible Clar formulas are a Clar sextet.
As a consequence, the sides of these \emph{non-Clar} hexagons have a net bond order of 5/4 and therefore are longer than the bonds sticking out of them (bond order 3/2). 
The atomic coordinates from DFT calculations second this reasoning as illustrated in the first subfigures in \ref{fig:2CF}a-c. 
All ribbons of this subclass show a $(\sqrt{3}\times\sqrt{3})\textrm{R}30^{\circ}$ pattern in which hexagons with longer sides (blue) are surrounded by a crown of slightly shorter bonds sticking out of them (dark red/black).

\subparagraph{Hexagon areas.}
The bond orders implied by the two Clar formulas of the subclass 2CF structures also
mean that the area of
the hexagons which do not house a Clar sextet in either Clar formula (six sides with bond order 5/4) should be larger
than the area of
the surrounding hexagons that are a Clar sextet in one of the two formulas (three sides with bond order 5/4 and three with bond order 3/2).
The DFT coordinates are in agreement with this as shown in the second subfigures in \ref{fig:2CF}a-c.
In a $(\sqrt{3}\times\sqrt{3})\textrm{R}30^{\circ}$ pattern, the non-Clar hexagons appear in dark gray, corresponding to a larger area, and are surrounded by the smaller hexagons, in light gray, which are Clar sextets in one of the two possible Clar formulas.

\subparagraph{Simulated STM images.}
As the $\pi$-electrons of the carbon atoms forming a non-Clar hexagon are in both Clar formulas engaged in forming a Clar sextet in one of the adjacent hexagons, never in the non-Clar hexagon itself, the local density of occupied $\pi$-states in these non-Clar hexagons should be expected to be lower than the ones of the neighboring hexagons. This interpretation is in very good agreement with the pattern of dark spots in simulated STM images with negative bias. These dark areas are located at the positions of the non-Clar hexagons (see the fifth subfigures in \ref{fig:2CF}a-c). The bright contributions reflect the higher density of occupied $\pi$-states in the bonds sticking out of the non-Clar hexagons. In positive bias STM images on the other hand, bright rings located at the non-Clar hexagons reflect the high density of unoccupied $\pi$-states in the corresponding region, again in perfect agreement with what one would expect from the Clar formulas. Note that the STM images with a given bias sign of class 2CF structures resemble very much the images with opposite sign bias of class 1CF structures. So, changing between the classes 1CF and 2CF has the same effect as staying in one class but changing the sign of the bias voltage.

\begin{figure}
	\begin{flushleft}
	\includegraphics[scale=1]{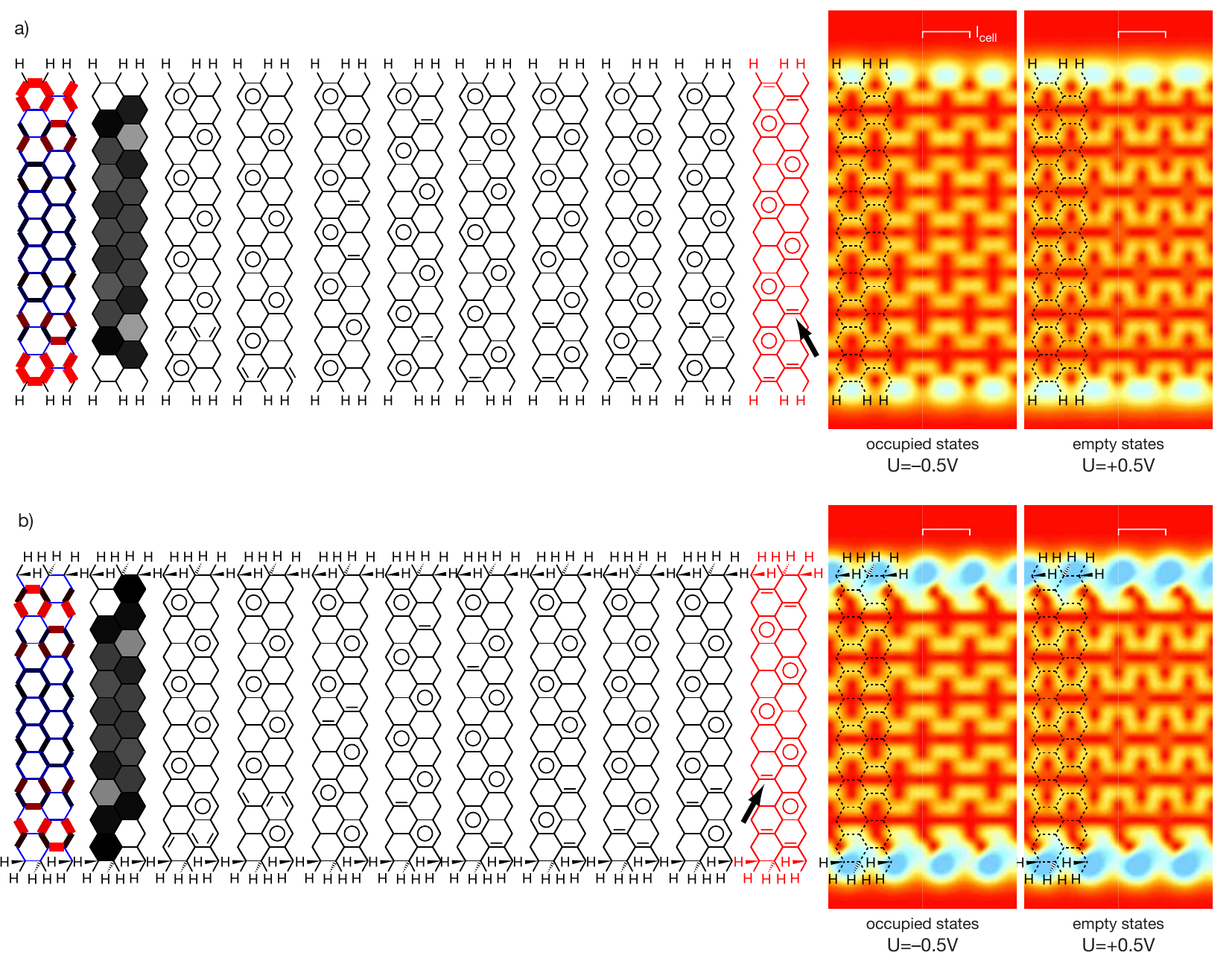}
  	\caption{Graphene ribbons with more than two equivalent Clar formulas, subclass nCF: ac(11) ribbon with width index $w=21$ (a) and ac(22) ribbon with $w=20$ (b). 
	In the negative bias STM images, every third horizontal bond remains dark.
	These bonds cannot be double bonds in a Clar formula with maximal number of Clar sextets.
	See the Clar representations to the right in red and the bonds indicated with arrows.
	The red configurations feature only six Clar sextets in (a) and five in (b), which is one less than in the other Clar formulas of the corresponding ribbons.
	}
	\label{fig:nCF}
	\end{flushleft}
\end{figure}

\paragraph{Subclass nCF: More than two Clar formulas}
The armchair ribbons not discussed so far, ac(11) with $w=3n,\; n$ being integer, and ac(22) with $w=(3n+2)$ belong to the subclass nCF, meaning that they have more than two equivalent Clar representations maximizing the number of Clar sextets. 
The Clar formulas of class nCF structures are not necessarily connected by symmetry operations among each other as one can see in \ref{fig:nCF}a and \ref{fig:nCF}b. Note that not all possible Clar formulas are shown for reason of space. 

\subparagraph{Bond lengths.} 
Except for the hexagons at the very edges of the ac(22) ribbons, all hexagons in the depicted examples become a Clar sextet in one or more of the possible Clar formulas. 
The $\pi$-electrons that are not part of a Clar sextet can either form horizontal double bonds parallel to the edges of the ribbons or 
double bonds in the lateral direction. 
When we have a closer look at the Clar formulas, we notice however, that every third horizontal bond cannot be a double bond in a Clar representation with maximal number of Clar sextets (see the last Clar representations to the very right, in red, in \ref{fig:nCF}a and \ref{fig:nCF}b).
If we consider a Clar representation in which such a bond is a double bond, the number of Clar sextets is necessarily lower than maximally possible. This leads to a slightly lower bond order of every third horizontal bond parallel to the edges of the ribbons. 
The DFT coordinates, however, indicate that 
the effect is of minor importance 
in the superposition of all the many possible Clar formulas and compared to the implications of the alternated forces on the edge atoms. 
The first subpicutres of \ref{fig:nCF}a and \ref{fig:nCF}b show that the effects of the different Clar formulas have cancelled out. 
Significant bond length alternations appear only at the very edges. 
In the interior of the ribbons, the bonds approach the graphene limit very fast and bond length alternations disappear.

\subparagraph{Hexagon areas.}
Also the effect a single Clar formula might have on the hexagon areas is evened out in the
superposition of the many possible Clar formulas. 
The DFT coordinates show an analogous picture as depicted in the second subfigures in \ref{fig:nCF}a and \ref{fig:nCF}b.
Significant area alternations appear only at the edges and can be attributed to changed effective forces on the edge atoms. 
In the interior of the ribbons, the hexagon areas attain a uniform limit.
This is in contrast to the class 1CF and 2CF cases where
patterns of hexagons with alternating areas
run through the whole width of the ribbons.

\subparagraph{Simulated STM images.}
Interestingly, the simulated STM images in \ref{fig:nCF} do not show a uniform, evened out signal as 
one might expect form the superposition of all possible Clar formulas. 
Instead, the negative bias STM images accentuate
a pattern with two out of three horizontal double bonds appearing bright and one dark.
The horizontal bonds appearing dark at negative bias correspond to bonds that cannot be double bonds in a Clar representation with maximal number of Clar sextets as pointed out before. So, the 
bond order inhomogeneity among the horizontal bonds,
implicated by the Clar formulas for this subclass,
has no remarkable influence on the geometry
but the alternating $\pi$-electron content of the corresponding bonds still appears in simulated STM images.

\subsubsection{Non-benzenoid ribbons}
Among the configurations considered in this study, the zz(1) and zz(2) ribbons make up the class of non-benzenoid ribbons. 

\begin{figure}
	\begin{flushleft}
	\includegraphics[scale=1]{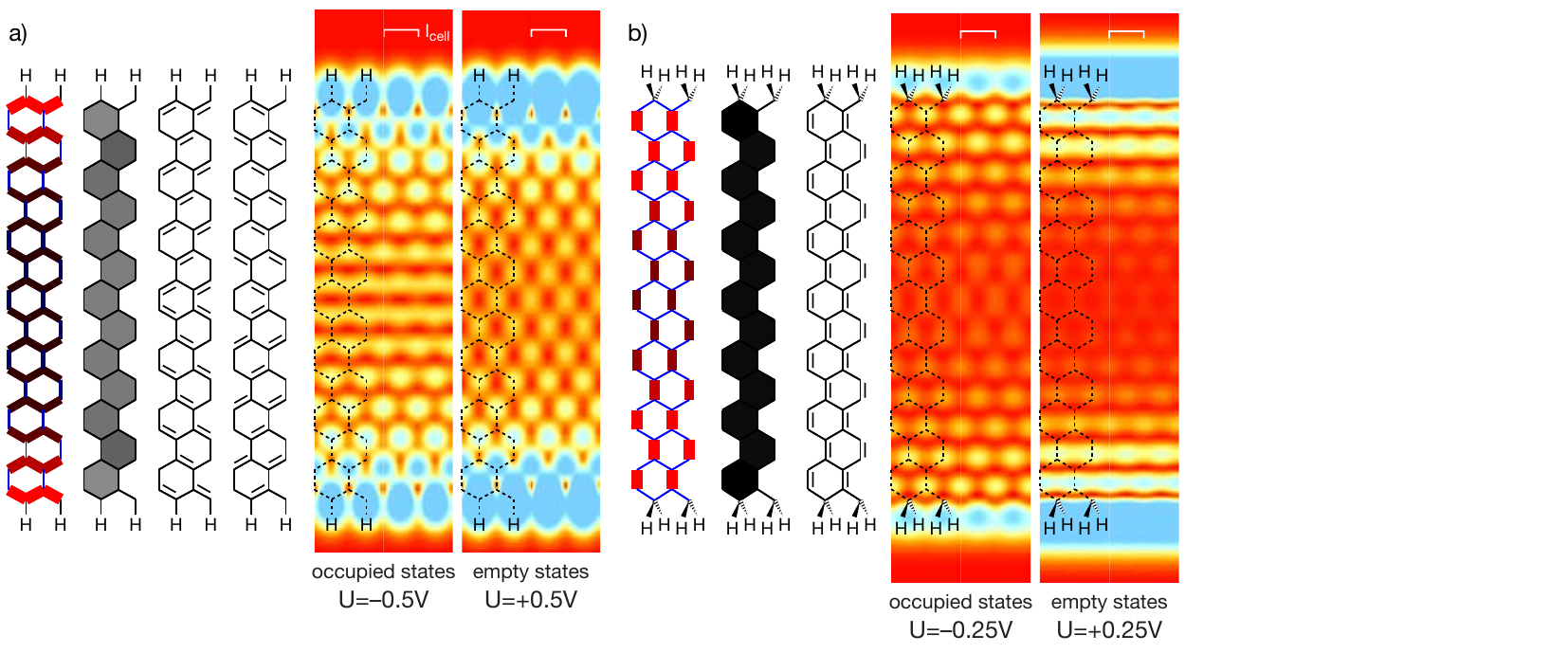}
  	\caption{Non-benzenoid graphene ribbons: (a) zz(1) and (b) zz(2), both with width index $w=13$. Both configurations feature a spin polarized edge state. In the zz(2) case, a lower bias voltage than $\pm0.5\,\textrm{eV}$ was chosen as otherwise the edge states would dominate the image completely.}
	\label{fig:NB}
	\end{flushleft}
\end{figure}

\subparagraph{Clar formulas.}
Under the premise that all four valence electrons of each carbon atom are engaged in bonds with neighboring atoms and no free radicals or dangling bonds are present,
the formation of Clar sextets in
the Clar representation of a zz(2) ribbon
is completely suppressed.
Instead, all bonds in the lateral direction of the ribbon are forced to be double bonds (third subfigure in \ref{fig:NB}b). 
Obviously, in this case the number of double bonds in the Clar formula of the unit cell increases with the width of the ribbon.
Such structures, in which all hexagons have two double bonds, are called fully \emph{quinoidal}\cite{Randic2003} (or \emph{quinonoid}\cite{Kertesz2005}).

\begin{figure}
	\begin{flushleft}
	\includegraphics[scale=1]{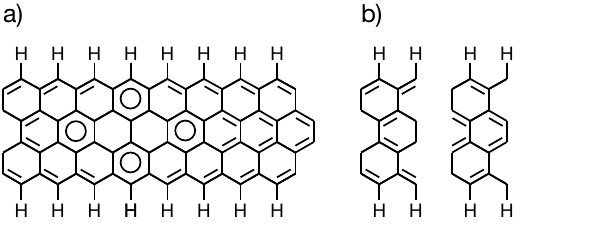}
  	\caption{
	Clar representation of a zz(1) ribbon with width index $w=3$ (a). Only a finite number of Clar sextets is allowed to form on the infinitely long ribbon. There is no unique choice for their position. As a consequence, the bonding is equivalently described by the superposition of two fully quinoidal Clar formulas (b).
	 }
	\label{fig:Clar_zz1}
	\end{flushleft}
\end{figure}

In the case of the zz(1) ribbons, the formation of Clar sextets is not forbidden in a strict sense. 
In a limited region of the infinitely long ribbon, Clar sextets can still form as illustrated in \ref{fig:Clar_zz1}a.\cite{Watson2001}
The Clar sextets, however, break the periodicity of the ribbon and there are infinitely many positions to place them on the infinitely long ribbon. 
The superposition of these (infinitely many) Clar representation with limited number of Clar sextets
is equivalent to the superposition of two fully quinoidal structures of the kind depicted in \ref{fig:Clar_zz1}b and \ref{fig:NB}a. 
Notice that in the quinoidal bond formulas, the bonds in the lateral direction of the ribbon are single bonds and double bonds are concentrated in the zigzag direction along the ribbon.

\begin{figure}
	\begin{flushleft}
	\includegraphics[scale=1]{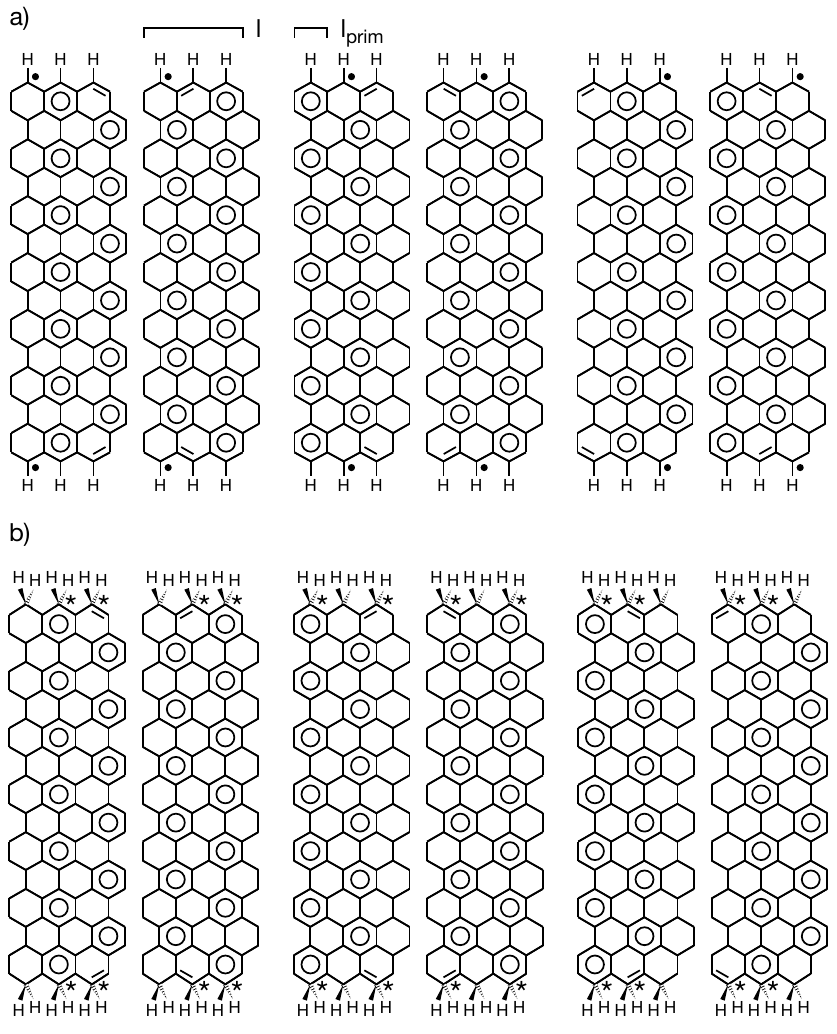}
  	\caption{
	Alternative Clar representations for the zz(1) (a) and the zz(2) case (b). At the cost of introducing unpaired electrons ($\bullet$) or holes ($\star$) at the edges, a (pseudo)-all-benzenoid $\pi$-electron distribution can be re-established.
	The periodicity $l$ of these representations is three times as long as the primitive periodicity for zigzag ribbons, $l_{\textrm{prim}}$.
	}
	\label{fig:NB_Clar_Radicals}
	\end{flushleft}
\end{figure}

By introducing radicals, i.e.\ unpaired electrons or holes, in the Clar representations, a (pseudo)-all-benzenoid $\pi$-electron distribution in the interior of the ribbons can be re-established for both non-benzenoid configurations (\ref{fig:NB_Clar_Radicals}).\cite{Mauri2008,Balaban2009}
Obviously, for ribbons wide enough, the gain in resonance energy favors this alternative (pseudo)-all-benzenoid bond configurations.
As the radicals represent uncompensated spin moments, the alternative Clar models also account for the spin polarized edge states found in DFT calculations on these kind of ribbons.\cite{Yamashiro2003, Son2006, Pisani2007} 
Interestingly, they hold even quantitatively: 
The Clar formulas depicted in \ref{fig:NB_Clar_Radicals} show one [two] radical[s] per edge and per supercell for zz(1) [zz(2)].
Hence, they imply a net spin moment of $\pm 1/3\,\mu_{\textrm{B}}$ per edge and per primitive unit cell for the zz(1) ribbon and one of $\pm 2/3\,\mu_{\textrm{B}}$ for zz(2). 
Integration of the DFT spin polarization, $\rho(\textrm{up})-\rho(\textrm{down})$, where $\rho(\textrm{up})$ and $\rho(\textrm{down})$ denote the spin-resolved electron densities, over half of the ribbon revealed a net spin moment of $\pm 0.30\,\mu_{\textrm{B}}$ on each side in the zz(1) configuration and of $\pm 0.63\,\mu_{\textrm{B}}$ in zz(2). 
The underestimation of the expected values by the DFT results may be seen as an indication
that the real (DFT) configuration is a mix of the fully quinoidal configuration of \ref{fig:NB} and the alternative (pseudo)-all-benzenoid configuration of \ref{fig:NB_Clar_Radicals}. 

The existence of unpaired electrons and holes at the edges of zz(1) ribbons
is also in agreement with results of an
investigation of the electronic ground state of higher acenes\cite{Bendikov2004,Jiang2008}.
There it was revealed that the ground state of acenes consisting of more than six fused benzene rings is an open-shell singlet state with antiferromagnetic order between the zigzag edges. It is not likely that this state is a diradical, since the net spin moment on the edges increases with the size of the acene.

\subparagraph{Bond lengths.}
The superposition of the alternative (pseudo)-all-benzenoid Clar formulas of \ref{fig:NB_Clar_Radicals} results in a complete equalization
of the bond orders in the interior of the ribbons and thus the implied $\pi$-electron distribution should lead to a uniform bond length throughout the ribbons.
In contrary to that, the DFT coordinates show 
a clear anisotropy of the bond lengths as illustrated in the first subfigures in \ref{fig:NB}a and \ref{fig:NB}b.
The zz(1) configuration exhibits significantly shorter bonds along the zigzag direction and longer bonds in the lateral direction of the ribbon whereas the zz(2) configuration shows short bonds in the lateral direction and long bonds in the zigzag direction.
This distribution of the bond lengths, however, is in perfect agreement with the localization of the double bonds in the fully quinoidal Clar representations depicted in \ref{fig:NB} and it might be seen as another
indication that the true DFT state is a mix of the fully quinoidal and the alternative (pseudo)-all-benzenoid configurations.

\subparagraph{Hexagon areas.}
For the hexagon areas, the 
fully quinoidal and the alternative (pseudo)-all-benzenoid Clar representation
 make no difference.
In both models, all hexagons in the interior of the ribbons
have the same bond configuration and therefore are expected to be of the same size.
This is in agreement with the DFT results presented in the second subfigures in \ref{fig:NB}a and \ref{fig:NB}b.
Notice, however, that 
as four out of the six bonds constituting a hexagon in the zz(2) ribbon are 
clearly longer than in ideal graphene,
these hexagons are considerably larger than the ones in graphene. 

\subparagraph{Simulated STM images}
The alternative (pseudo)-all-benzenoid Clar representations imply a magnetic state localized at the edges of both, zz(1) and zz(2), ribbons. But besides that, they predict a uniform distribution of the $\pi$-electrons in the interior of the ribbons. 
For the zz(1) configuration, the fully quinoidal Clar representations, on the other hand, hint at a high density of occupied states in the zigzag bonds along the direction of the ribbon and a low density of occupied states in the bonds in lateral direction.

In fact, the most dominant feature in the simulated STM images of both ribbons in \ref{fig:NB}a and \ref{fig:NB}b is the signature of the edge state that decays towards the interior of the ribbons.
In addition to that, in the interior of the zz(1) ribbon, where the edge state has decayed, the negative bias STM image highlights horizontal structures corresponding to the double bonds in the zigzag direction along the ribbon. 
Complementary to that and in agreement with the fully quinoidal Clar representation, at positive bias, 
a high density of empty electronic states in the lateral bonds 
is reflected by 
bright contributions in the corresponding spots. 

\subsection{Ribbon width dependences}
\subsubsection{Band gap}
\begin{figure}
	\begin{flushleft}
	\includegraphics[scale=1]{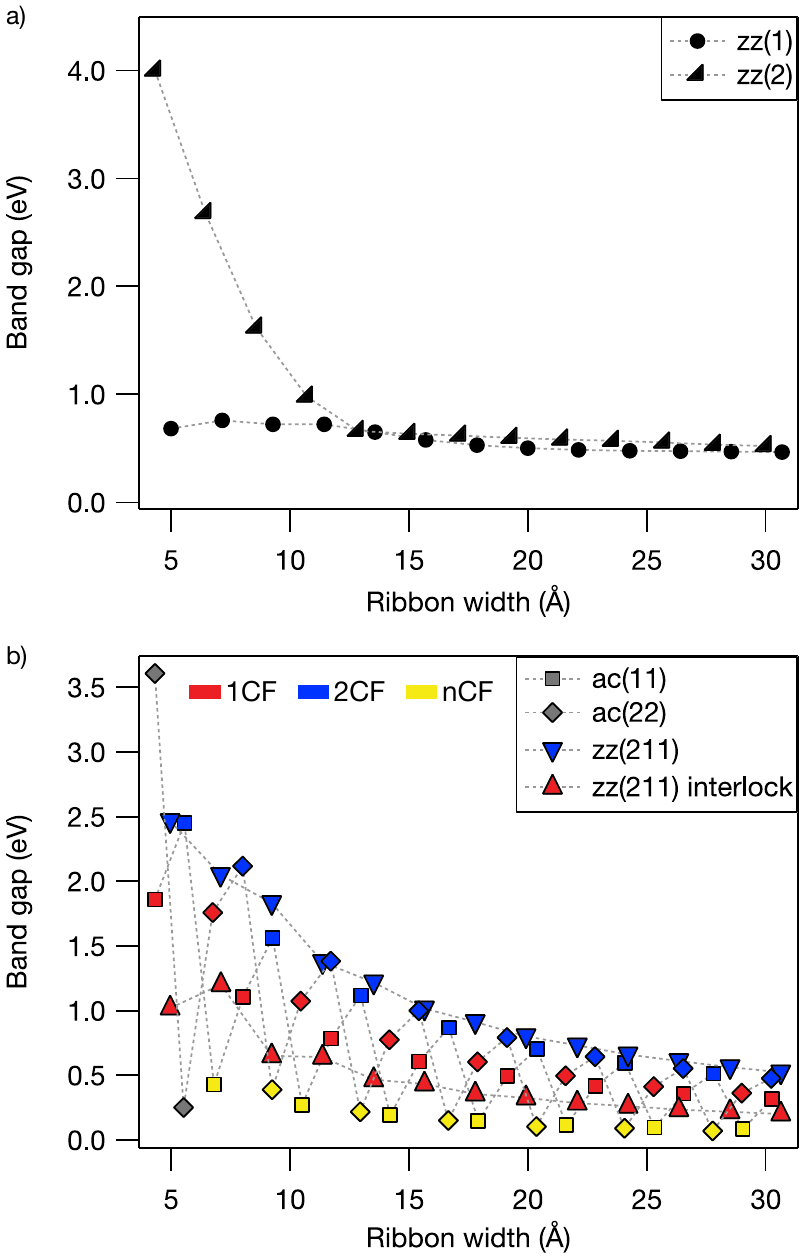}
  	\caption{Band gap versus ribbon width for non-benzenoid ribbons (a) and (pseudo)-all-benzenoid ribbons (b). The (pseudo)-all-benzenoid configurations are color-coded according to the subclass they belong to (see text).}
	\label{fig:bandgap_vs_width}
	\end{flushleft}
\end{figure}

Even though DFT-GGA is known to underestimate the band gaps, we would still like to include a qualitative discussion of the width dependence of this quantity. \ref{fig:bandgap_vs_width} shows the band gap versus the width of the ribbons considered in this study. 
While except for the smallest four zz(2) ribbons\footnote{According to our calculations, no magnetic edge state is present in the narrowest four zz(2) ribbons.}
the band gaps of the non-benzenoid configurations (\ref{fig:bandgap_vs_width}a) are almost equal and nearly independent of the width,
the (pseudo)-all-benzenoid ribbons (\ref{fig:bandgap_vs_width}b) show rich characteristics. 
The threefold periodicity of the ac(11) band gap is a well-known phenomena.\cite{Ezawa2006, Barone2006, Son2006b,Okada2008} 
Here we report the same kind of oscillations also for ac(22) ribbons. 
One can see in \ref{fig:bandgap_vs_width}b that among the ac(11) and ac(22) ribbons, class 2CF structures represent band gap maxima and class nCF structures band gap minima whereas 
class 1CF configurations have intermediate band gaps. 
The band gap minima of class nCF structures coincide with the zero band gap of two-dimensional graphene insofar as according to our classification scheme, two-dimensional graphene would also fall into class nCF, as it has three equivalent Clar formulas. 
Interestingly, the band gaps of the zz(211) and zz(211)-interlock ribbons line up with the values of the other structures in their corresponding class, too. The result is a picture in which the band gaps are separated according to the three subclasses,
with class 1CF band gaps being enveloped by class 2CF maxima and class nCF minima. 
Small band gaps can serve as indication for higher chemical reactivity.\cite{Aihara1999}

\subsubsection{Edge energy}
\begin{figure}
	\begin{flushleft}
	\includegraphics[scale=1]{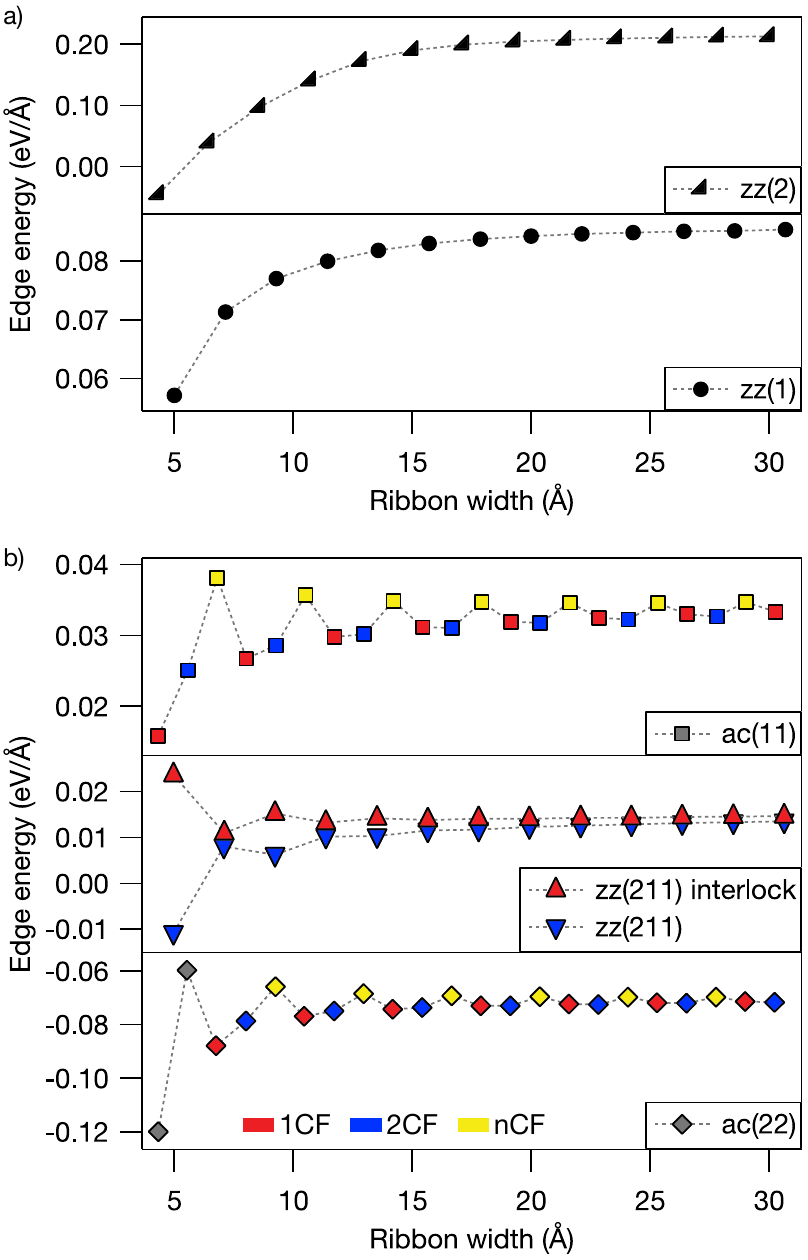}
  	\caption{Edge energy versus ribbon width for non-benzenoid ribbons (a) and (pseudo)-all-benzenoid ribbons (b). The (pseudo)-all-benzenoid configurations are color-coded according to the subclass they belong to (see text).}
	\label{fig:eedge_vs_width}
	\end{flushleft}
\end{figure}

\ref{fig:eedge_vs_width} shows the edge energy with respect to hydrogen gas, 
\begin{equation}\label{eq:Eedge}
  \textrm{E}^{\textrm{edge}}_{\textrm{H}_2} =
  \frac{1}{2\,l_{\textrm{cell}}} \, \left(
  \textrm{E}_{\textrm{ribbon}} -
  \textrm{n}_{\textrm{C}}\,\textrm{E}_{\textrm{C}} -
  \textrm{n}_{\textrm{H}}\,\frac{1}{2}\,\textrm{E}_{\textrm{H}_{2}}
   \right) \ ,
\end{equation}
as a function of the width. 
This quantity reflects the energy necessary for the formation of the edges.
In the above definition, $\textrm{n}_{\textrm{C}}$ ($\textrm{n}_{\textrm{H}}$) denotes the number of carbon
(hydrogen) atoms in the unit cell of the ribbon, 
$\textrm{E}_{\textrm{ribbon}}$, $\textrm{E}_{\textrm{C}}$,
and $\textrm{E}_{\textrm{H}_2}$ stand for the total DFT energies of the ribbon, one
carbon atom in bulk graphene, and an isolated hydrogen molecule, respectively.

The edge energy of the non-benzenoid configurations zz(1) and zz(2) increases monotonously and converges to about $85\,\textrm{meV/\AA}$ for zz(1) and to $215\,\textrm{meV/\AA}$ for zz(2) (see \ref{fig:eedge_vs_width}a). 
Also the values for the (pseudo)-all-benzenoid zz(211) and zz(211)-interlock ribbons converge mainly monotonously to a common asymptotic limit of $14\,\textrm{meV}$ (\ref{fig:eedge_vs_width}b). 
The two armchair configurations, ac(11) and ac(22), on the other hand, show an oscillating convergence with a threefold periodicity.
In these oscillations, class nCF structures constitute maxima and class 1CF structures minima, at least in narrow ribbons. 
For broader ribbons, class 1CF and class 2CF structures have degenerate values, but class nCF values are still slightly larger.
The asymptotic limit for the ac(11) configurations lies at about $33\,\textrm{meV/\AA}$ and the one for the ac(22) configurations at $-72\,\textrm{meV/\AA}$. 
Other studies discussing the edge energy oscillations of single-hydrogen-terminated or undecorated armchair ribbons include Refs.~\cite{Kawai2000,Okada2008}.

For the ac(22) ribbons, separate calculations in the ac(22)-mirror and ac(22)-inversion configuration (see \ref{fig:notation}) have been performed.
The configurations with mirror symmetry exhibit a lower total energy than the one with inversion symmetry for width indices $w=1,6,7,8$. 
In the other cases with $w<10$, the ac(22)-inversion configuration is favored.
For ribbons with $w\ge10$, i.e.\ ribbons with a width of at least $15.4\,\textrm{\AA}$, the two configurations are energetically equivalent.

\subsubsection{Edge strain}
\begin{figure}
	\begin{flushleft}
	\includegraphics[scale=1]{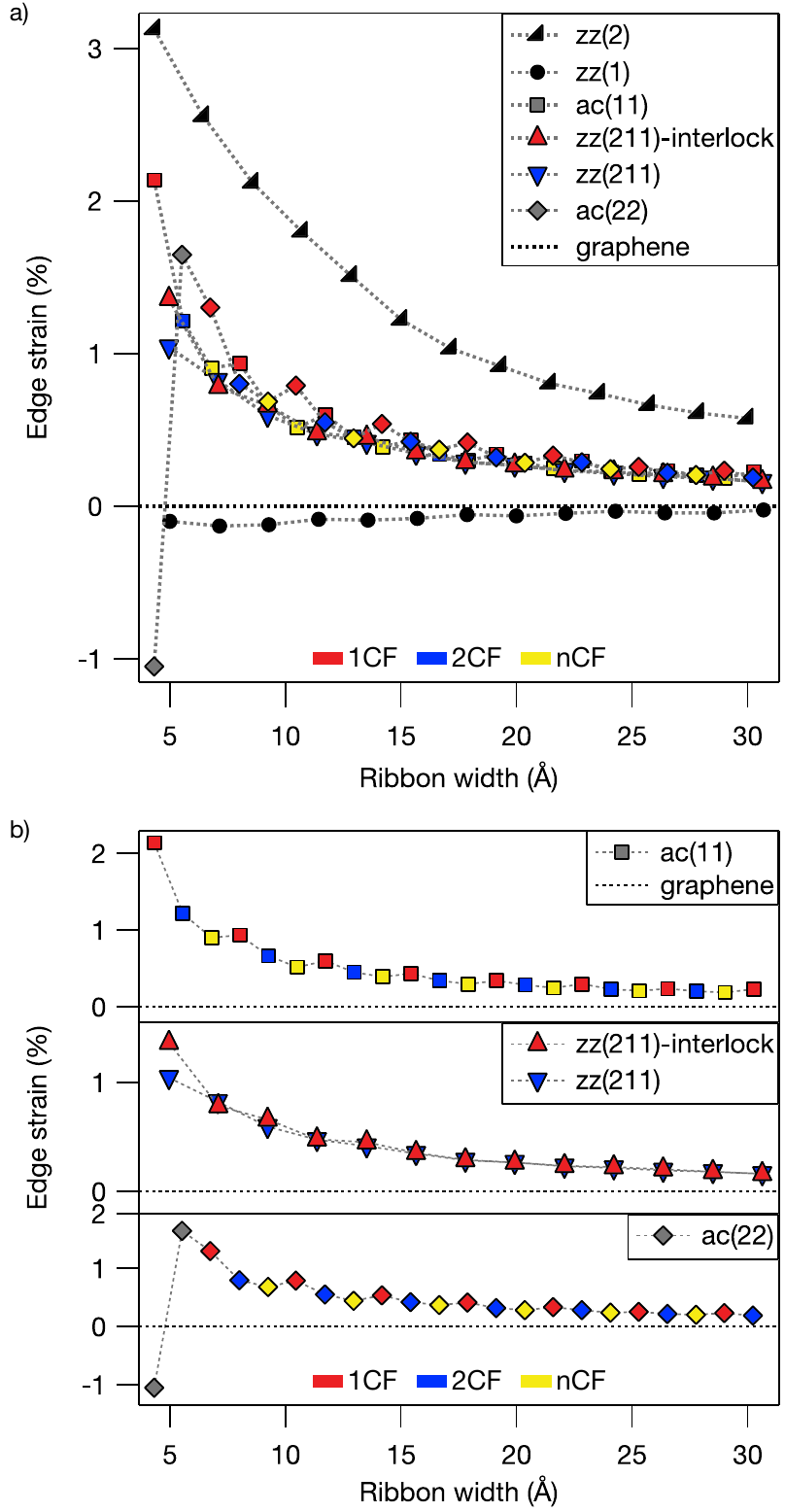}
  	\caption{Edge strain as a function of the width for all ribbons considered in this study (a) and for the (pseudo)-all-benzenoid ribbons in detail (b). The (pseudo)-all-benzenoid configurations are color-coded according to the subclass they belong to (see text). 
	}
	\label{fig:alat_vs_width}
	\end{flushleft}
\end{figure}

The edge-induced strain describes the variation of the equilibrium lattice length of the ribbons with respect to the one of two-dimensional graphene.
For ideal graphene, the lattice constant found with our computational setup amounts to $2.47\,\textrm{\AA}$ which compares well to the experimental lattice constant of graphite, $2.46\,\textrm{\AA}$.\cite{Trucano1975}
\ref{fig:alat_vs_width} shows this parameter as a function of the ribbon width.
With increasing width, the lattice length of all ribbons converges to the one of two-dimensional graphene. 
Except for the zz(1) case, the value of all configurations converges from above. 
The edge strain of the (pseudo)-all-benzenoid ribbons ac(11), zz(211), zz(211)-interlock, and ac(22) lies between the extrema represented by the two non-benzenoid configurations zz(1) and zz(2). 
This is in agreement with the fully quinoidal Clar formulas and the bond length analysis in \ref{fig:NB}. 
There we found a preference for double bonds along the periodic zigzag direction for zz(1) ribbons, and hence we would expect the lattice length in this direction to be smaller than in two-dimensional graphene. 
In the zz(2) configurations, on the other hand, the bonds along the periodic zigzag direction are forced to be single bonds in the radical-free Clar formula, making them longer than in graphene. 
And in addition, the repulsion between the edge hydrogen atoms also acts in this direction, leading to an enlarged lattice length.
Interestingly, the edge strain of the armchair configurations again shows a small oscillation with a periodicity of three, in line with the alternation between the three (pseudo)-all-benzenoid subclasses as shown in \ref{fig:alat_vs_width}b.
Maxima correspond to class 1CF structures, minima to class nCF configurations. 

\section{Discussion}
Our simulated STM images represent an idealization as for instance effects of the substrate and the physics of the tip were not considered.
However, STM studies differentiating Clar sextets form non-Clar hexagons are not bound to computer simulations. 
Experimental observations
were achieved in PAHs,\cite{Ohtani1988, Samori2002, Gunther2005, Mullen2007} on the edges of graphite sheets,\cite{Kobayashi2006} and even in graphene nanoribbons (supplementary informations of Ref.~\cite{Tapaszto2008}).

In particular, we would like to point out that in Ref.~\cite{Kobayashi2006}
ring patterns near zigzag edges were recorded 
at positive sample bias which match very well with our simulations on the zz(211) ribbons shown in \ref{fig:2CF}c.
Near armchair edges, the STM patterns reported in Ref.~\cite{Kobayashi2006} resemble very much our results for subclass nCF armchair ribbons presented in \ref{fig:nCF}. 
See Fig.\,S1 in the supporting informations.
However, one has to be cautious when comparing the results of Ref.~\cite{Kobayashi2006} to our simulations: In Ref.~\cite{Kobayashi2006} the edges of large graphite sheets were investigated whereas here we considered comparably narrow graphene ribbons whose edges still interact.

An STM image of a 10 nm wide graphene ribbon claimed to have armchair edges was presented in Figure S3 (right) in the supplementary material of Ref.~\cite{Tapaszto2008}. This image, recorded at $U=0.2\, \textrm{V}$, shows in some parts a pattern of bright rings that compares very nicely to our simulations of the zz(211) ribbon at positive bias (\ref{fig:2CF}c), indicating that the observed ribbon probably runs in zigzag direction.
See Fig.\,S2 in the supporting informations.

Despite these encouraging experimental results,
an atomically resolved STM characterization of smooth-edged graphene nanoribbons still remains to be achieved. In our simulations we showed that the edges dominate the STM patterns far into the interior of the ribbons. These patterns can give valuable informations on the edge configuration even if an atomic resolution is not achieved at the very edges.
Furthermore, we remark that in STM images on multilayer graphene or graphite, only the atoms corresponding to one of the two graphene sublattices are visible because of the AB stacking. 
The $(\sqrt{3}\times\sqrt{3})\textrm{R}30^{\circ}$ diamond and 
honeycomb superstructures observed in Ref.~\cite{Niimi2005, Niimi2006} can partially be understood by a corresponding sublattice modulation of our results.

Besides STM, helpful information for the characterization of graphene ribbons is also expected from
Raman spectroscopy.
The Raman spectrum measured in the middle of a perfect graphene crystal features a G-peak at around $1580\,\textrm{cm}^{-1}$.
In the neighborhood of lattice defects or near the edges, STM images indicated a $(\sqrt{3}\times\sqrt{3})\textrm{R}30^{\circ}$ superstructure of the
charge density in which every third hexagon stands out.\cite{Niimi2005, Niimi2006,Niimi2006b,Mallet2007,Brihuega2008}
In the supercell accounting for this reconstruction, the K-phonons are folded back to $\Gamma$ and activate a D-peak at around $1350\,\textrm{cm}^{-1}$.
This D-peak can be used as a signature of defects in graphene.\cite{Ferrari2006, Chen2009, Ni2009}
The phonon near K responsible for the D-peak represents a shrinking and expanding of every third hexagon in accordance with the $(\sqrt{3}\times\sqrt{3})\textrm{R}30^{\circ}$ reconstruction of the charge density.

In the case of graphene edges, it was claimed that only armchair edges could induce a D-line peak, suggesting that the presence of such a peak can be used as evidence for either armchair edges or defects.\cite{Pimenta2004,You2008}
What has not been considered so far is the possibility of a zigzag reconstruction that induces the same kind of charge density and geometry reconstruction as at armchair edges or near impurities. 
Here we have shown that the zz(211) and zz(211)-interlock ribbons (subclasses 1CF and 2CF) do show such a superstructure and in addition have a band structure that folds the K-point back to $\Gamma$. They thus should also be expected to produce a D-line peak.

\section{Conclusions}
In this article, we have demonstrated the implications Clar's theory of the aromatic sextet has on a set of hydrogen-terminated graphene ribbons. Depending on whether the major portion of the $\pi$-electrons is engaged in Clar sextets or not, graphene ribbons can be distinguished into two fundamental classes, here called (pseudo)-all-benzenoid and non-benzenoid. For the (pseudo)-all-benzenoid ribbons we have proposed a further distinction according to the number of equivalent Clar formulas they have. 
This way, ribbons with similar edge-induced Clar representations are grouped independently of their edge geometry.

An investigation of DFT based atomic coordinates and simulated STM images has revealed that the Clar formulas correctly account
for the $\pi$-electron distribution and geometric aspects such as bond length and hexagon area alternations.
The ribbons in each group of the proposed classification scheme share 
a common STM signature that extends far into the interior region, reflecting the edge-induced $\pi$-electron distribution in agreement with the corresponding Clar formulas.
We have pointed out that the presence of a Raman D-line peak alone should not be used as indication for armchair edges since there exist also zigzag reconstructions expected to activate the corresponding phonons.

By introducing radicals at the edges of the non-benzenoid ribbons,
a bond configuration can be restored that includes nearly all $\pi$-electrons in Clar sextets.
As this alternative bond configuration increases the resonance energy, at large enough dimensions, it becomes energetically favored.
The notation of radicals at the edges is in line with the spin polarized edge states found in DFT calculations on these ribbons. 
We have shown that there is even a quantitative agreement in the net magnetization of these alternative bond configurations and our DFT results.

Finally, we have found a correlation to our classification scheme also in a series of calculations investigating the width dependence of the electronic band gap, the edge energy, and the edge-induced strain. 

These findings may play an important role on the way to experimental structure characterization. They emphasize also that Clar's theory is an intuitive and powerful tool covering many aspects of sp$^2$ bonded carbon materials such as graphene ribbons.


\suppinfo
Comparison of our theoretical results with measurements provided in Ref.~\cite{Kobayashi2006} and ~\cite{Tapaszto2008} as well as complete lists of authors for Ref.~\cite{Delgado2008} and ~\cite{Giannozzi2009} .
This information is available free of charge via the Internet at http://pubs.acs.org/.


\bibliography{Manuscript-Wassmann+Mauri-condmat}

\end{document}